# An SERS study of the galvanostatic sequence employed for the electrochemical deposition of Copper in the fabrication of Interconnects

Lucia D'Urzo* and Benedetto Bozzini

Dipartimento di Ingegneria dell'Innovazione,

Università del Salento (formerly Università di Lecce), via Monteroni, I-73100 Lecce, Italy

* Corresponding author: lucia.durzo@unile.it

ABSTRACT

This paper reports the first study carried out by surface-enhanced Raman spectroscopy (SERS) during the galvanostatic electrodeposition (ECD) of copper from an acidic sulphate solution, in the presence of polyethylene glycol (PEG), bis-(3-sulfopropyl)-disulfide Na salt (SPS), benzyl-phenyl modified polyethyleneimine (BPPEI) and chloride ions. The analysis of SERS spectra recorded during electrodeposition allowed to get an insight into the complex interfacial behaviour of the organic blend, in terms of co-adsorption and reactivity. At open-circuit (OC), the additives co-adsorb on the copper cathode. Upon increasing the cathodic polarization, progressive SPS-scavenging action of PEG was observed. BPPEI is adsorbed in the entire process window and cathodic reaction products of PEG were identified. The joint action of the organic additives yields a continuous deposit with crystallites of submicron dimensions, as revealed by Scanning Electron Microscopy (SEM).

## 1 INTRODUCTION

Electroplated copper is becoming the electric contact material of choice in the semiconductor industry, because it offers lower line resistance and better electromigration performance compared to conventional Al and Al-alloy metallization [1]. As a consequence of the interaction of different organic additives and of special electrochemical polarisation schemes “superfilling” behaviour can be achieved, yielding void- and seam-free electrodeposits in high-aspect ratio features. The organic additives in acid Cu plating baths are commonly categorised as [2, 3, 4, 5]: (i) carriers, (ii) brighteners or accelerators and (iii) levellers. Historically, Polyethylene glycol (PEG) of several molecular weights (MWs) [6-11] and Bis-(3-sulfopropyl)-disulfide Na salt (SPS) [12-30] were used in the Cu ECD as suppressor and accelerator, respectively. Nowadays, the selection of the levelling agent represents one of the crucial factors for an effective transition of Cu ECD towards the most advanced technology nodes. A novel polymeric leveller based on polyvinilpyrrollidine has been recently presented in [31]. In this paper we deal with copper electrodeposition from an acidic sulphate bath in the presence of chloride ions, PEG, SPS, and Benzyl-phenyl modified polyethyleneimine (BPPEI). The choice of the latter is based on our previous experience with benzyl-dimethyl-phenyl ammonium ($BDMPA^+$) $[Ph\text{-}N(CH_2)_2\text{-}CH_2\text{-}Ph]^+$ as an Au-plating additive [32-34]: we employed a functionalised polyethylene-imine type polymer containing the same quaternary ammonium functionality in the backbone as a model grain refiner for Cu plating [35÷38]. The molecular scheme of this polymer is: $\text{-}[CH_2\text{-}CH_2\text{-}N(Ph,CH_2\text{-}Ph)^+]_x\text{-}[CH_2\text{-}CH_2\text{-}N(CH_2\text{-}CH_2\text{-}NH_2)_z]_y\text{-}[CH_2\text{-}CH_2\text{-}NH]_t\text{-}$

Our electrochemical study is based on SERS, complemented by morphological SEM investigations of the effects of organic additives on the electroplated copper. To the best of our knowledge, many SERS studies related to single-additive containing bath have been published in [39-41]; however the behaviour of the entire blend, that is ultimately responsible for the quality of the coating, was not addressed by in situ spectroscopic study until now.

## 2 EXPERIMENTAL

The composition of the electrodeposition bath was: $CuSO_4 \cdot 5H_2O$ (Sigma Aldrich, Germany) 20mM, $H_2SO_4$ 0.5M. To this solution we added: NaCl (Sigma Aldrich, Germany) 50 ppm, PEG MW 1500 (Fluka, Germany) 300 ppm, SPS (Raschig GmbH, Germany) 6 ppm and BPPEI specifically synthesized for this research, 10 mg $L^{-1}$. SERS measurements were performed with a LabRam confocal Raman system. Excitation at 633 nm was provided by a 12 mW He-Ne laser. A 50× long-working-distance objective was used. In situ spectroelectrochemistry was carried out in a cell with a vertical polycrystalline Cu disc working electrode of diameter 5 mm embedded in a Teflon cylindrical holder. The counter electrode was a Cu cylinder, coaxial with the working electrode holder. An Ag/AgCl (KCl 3M) reference electrode was used, placed in a separate compartment; potentials are reported on the Ag/AgCl scale. The spectra acquisition time ranged between 30 and 60 s. The morphology of the samples was studied with a Cambridge Stereoscan 360 SEM. The electron source was $LaB_6$.

## 3 IN SITU SURFACE ENHANCED RAMAN SPECTROSCOPY

SERS spectra were recorded at OC and 1, 2, 5, 6, 8 and 10 mA $cm^{-2}$. This range of c.d.s corresponds to that typically adopted in industrial processes, implementing galvanostatic electrodeposition or sequences of c.d. ramps [42]. In Figure 1 we show a typical spectrum obtained at open-circuit (0.056 mV). In this figure, the following features related to PEG were found:

i) Weak bands corresponding to the methylene group vibrations at: 1220/1270 $cm^{-1}$ ($CH_2$ twisting) and 1420 $cm^{-1}$ ($CH_2$ scissoring) [43];

ii) Methylene asymmetric (2925/2970 $cm^{-1}$) stretching [43];

iii) The weak broad band at 3070 $cm^{-1}$ has been assigned to the stretching of the C-H bond in the RHC= group, owning the PEG fragmentation at the cathode during a multi step process that yields moieties containing vinyl ether group, as previously reported in [44, 45, 49].

As far as SPS is concerned, the following diagnostic features were identified [46, 47, 48]:

i) 630 $cm^{-1}$ assigned to C-S stretching;

ii) 695/735/798 $cm^{-1}$, assigned to $SCH_2$ rocking

iii) 1040 $cm^{-1}$, assigned to the stretching of the sulfonic group.

Except for the strong band at 1005 $cm^{-1}$, assigned to an aromatic ring stretching mode [32, 33, 37, 43], the identification of BPPEI features was in some cases complicated by the relatively poor signal intensity and some overlapping with $CH_2$ twisting and scissoring of PEG. Nevertheless, the stretching of the dimethyl aniline ring at 1600 $cm^{-1}$ and the Ph-N stretching of quaternary ammonium at 1340 $cm^{-1}$ can be unambiguously identified. The band at 1460 $cm^{-1}$, assigned to semicircle stretching mixed with a C-H bending [43] mode, is better resolved at high current densities (see below and [38]).

The peak observed at 294 $cm^{-1}$ was assigned to Cu-O stretching [46, 47, 48] , even though we are aware that Cu-O readily dissolves in acidic environment, nevertheless it is well known that surface-enhancement effects might be responsive to trace amounts of possibly transiently adsorbed species.

A list of the observed Raman bands, their indicative intensities and their assignments are provided in Table 1. When the bath was cathodically polarized at 1 $mA/cm^2$, we observed (Figure 2):

i) A notable reduction of the SERS signal intensity, probably due to the formation of a fresh, but SERS-inactive copper layer;

ii) The intensity of the Cu-O stretching decreases, due to fresh copper deposition;

iii) Three new bands appeared at 425 $cm^{-1}$, 530 $cm^{-1}$ and 850 $cm^{-1}$, assigned respectively to C-S bending, SPS disulfide bridge stretching and to the $SCH_2$ deformations modes;

iv) The relative intensity of all other SPS bands increased;

v) The intensities of PEG-related $CH_2$ stretching (2925/2970 $cm^{-1}$) and vinyl-related CH stretching (3070 $cm^{-1}$) tend to decrease or disappear [44, 45, 49].

Further spectral changes brought about by shifting the current from 1 to 10 mA cm$^{-2}$, can be summarised as follows (Figures 2-4):

i) For Raman shifts lower than 1000 cm$^{-1}$, strong SPS-related features show the adsorption of the salt in the entire cathodic window (Figure 2). This result is consistent with our previous potentiostatic experiments, carried out in an acidic copper sulphate bath containing SPS in the presence and in the absence of $Cl^{-}$ [47, 48]. As far as PEG is concerned, its typical carbonyl asymmetric stretching (at ~ 800 cm$^{-1}$) cannot be detected, due to its possible overlap with the stronger $SCH_2$ deformation modes. At 10 mA cm$^{-2}$ a new small feature appears at 960 cm$^{-1}$, that can be assigned to the BPPEI out of plate CH wagging [43] (Figure 2).

ii) For Raman shifts in the interval 1000÷2800 cm$^{-1}$ the most prominent peak was found at ~1040 cm$^{-1}$. All the features related to the methylene deformation of PEG exhibit a very low intensity that tends to decrease with increasing cathodic polarisation. BPPEI bands can be observed, with weak or medium intensity (Figure 2). The peak at 1040 cm$^{-1}$ was found to be a convolution of 3 peaks (Figure 3), respectively assigned to: ring stretching mode of BPPEI aromatic groups (~1005 cm$^{-1}$), vibration of PEG methylene groups (1020 cm$^{-1}$ [44, 45, 49]) and stretching of SPS sulfonic termination (1040 cm$^{-1}$). The relative intensities of these peaks change with the polarization. In particular, the ratio between the intensities of PEG- and SPS-related features tend to decrease as the c.d. increases, while the BPPEI band is better defined at higher c.d. values.

iii) For Raman shifts higher than 2800 cm$^{-1}$, the band assigned to the vinyl $CH_2$ stretching disappears (Figure 4). Overall, the intensity of methylene symmetric and asymmetric stretching bands decreases. No peaks related to SPS or BPPEI are present in this range.

The analysis of our SERS spectra allows to get an insight into the complex behaviour of this organic mixture during the electroplating process. At open circuit, SPS, PEG and BPPPEI bands clearly show the co-adsorption of the three additives on the electrode surface. Some moderate reactivity of PEG was also found coherently with the behaviour of baths containing PEG as the sole additive [44, 46]. The main reaction product has been identified in terms of polymer subunits containing the vinyl group R-CH=$CH_2$ as thoroughly discussed in [44, 45, 49]. When the cathodic polarisation is turned on, SPS features tend to dominate the spectra. We cannot identify any obvious changes in the spectral pattern that can demonstrate electrodic reaction of SPS. In particular, the stretching of the disulfide bridge was detectable at all c.d.s studied. These results are in good agreement with previous experiment carried out in the Copper electrodeposition bath containing SPS and $Cl^-$ only [48]. As far as PEG is concerned, the peak assigned to vinyl as a polymer breakdown product tends to disappear already at 1 mA $cm^{-2}$. This fact can be explained with its incorporation into the growing cathode, in correspondence of non-SERS-active sites [36] or by desorption. Moreover, the SPS-scavenging action of PEG has been reported in [2], based on electrochemical measurements. Finally, the adsorption of BPPEI by its aromatic moieties is proven in the entire cathodic range covered in this experiment. No changes in the spectral pattern, suggesting polymer reactivity, could be detected.

4 SCANNING ELECTRON MICROSCOPY

In Figures 5 and 6 we show in-plane SEM micrographs of thick (thickness in excess of 30 μm) Cu layers deposited galvanostatically at 2 mA $cm^{-2}$ from the following deposition baths:

i) $CuSO_4 \cdot 5H_2O$ 20mM, $H_2SO_4$ 0.5M, NaCl 50ppm, PEG MW 1500, 300 ppm, SPS 6 ppm and 3-Diethylamino-7-(4-dimethylaminophenylazo)-5-phenylphenazinium chloride (Janus Green B, JGB) 1 mg $L^{-1}$;

ii) $CuSO_4 \cdot 5H_2O$ 20mM, $H_2SO_4$ 0.5M, NaCl 50ppm, PEG MW 1500, 300 ppm, SPS 6 ppm and BPPEI 10 mg $L^{-1}$.

The former bath, containing a leveller widely studied in the literature [2, 38, 50], is commonly regarded as the benchmark for the evaluation of the grain refining power of innovative levellers. In the presence of JGB, a compact, grainy layer is obtained (Figure 5). Primary crystallites, deriving from nucleation events, seem to split during growth into sub-grain domains. Overall, JGB at this c.d. exhibits a slight grain-refining activity (average grain dimension ca. 1-3.5 μm). BPPEI, instead, shows a remarkable grain-refining effect. In the presence of the polymer, a continuous layer is obtained. Pictures taken at high magnification (3000X and 6000X) reveal the formation of elongated crystallites, of submicron dimensions (Figure 6).

5 CONCLUSIONS

The synergistic behavior of PEG, SPS and BPPEI during galvanostatic copper electrodeposition can be summarized as follows:

i) at open-circuit, accelerator, suppressor and leveller co-adsorb on the copper cathode.

ii) when the cathodic polarization is applied, the relative surface coverage between SPS and PEG increases with the polarization; BPPEI-related bands are present in the entire process window and look better defined at higher c.d.;

iii) the joint action of the organic additives yields a continuous deposit. Compared to the benchmark leveller JGB, BPPEI allowed the formation of crystallites of submicron dimensions.

To the best of the authors' knowledge, the simultaneous action of all the types of additives used for the electrodeposition of Cu interconnects during a galvanostatic process mimicking a state–of-art manufacturing process has been clarified at a molecular level for the first time in this research.

Acknowledgements

Highly qualified and continuous technical assistance with electrochemical experiments and SEM are kindly acknowledged to Francesco Bogani and Donato Cannoletta (Dipartimento di Ingegneria dell'Innovazione, Università di Lecce, Italy), respectively

<table>
<tr><th>OC</th><th>1<br>mA cm$^{-2}$</th><th>2<br>mA cm$^{-2}$</th><th>5<br>mA cm$^{-2}$</th><th>6<br>mA cm$^{-2}$</th><th>10<br>mA cm$^{-2}$</th><th>Assignment</th></tr>
<tr><td>--</td><td colspan="5">425 cm$^{-1}$, weak</td><td>C-S bending</td></tr>
<tr><td>--</td><td colspan="5">530  cm-1, weak</td><td>S-S stretching</td></tr>
<tr><td colspan="6">630 cm$^{-1}$, medium-strong</td><td>C-S stretching  and $SO_4^{=}$ bending band</td></tr>
<tr><td colspan="6">695/735/798 cm$^{-1}$, medium- strong</td><td>$SCH_2$ rocking</td></tr>
<tr><td></td><td colspan="5">850 cm$^{-1}$, strong</td><td>aromatic CH wag</td></tr>
<tr><td colspan="5"></td><td>960 cm$^{-1}$, weak</td><td>out of plate C-H wagging [43]</td></tr>
<tr><td colspan="6">1005 cm$^{-1}$, strong</td><td>Ring stretching mode</td></tr>
<tr><td colspan="6">1040 cm$^{-1}$</td><td>$SO_3^{-}$ symmetric stretching</td></tr>
<tr><td colspan="6">1220/1270 cm$^{-1}$, weak to very weak</td><td>$CH_2$ wagging and twisting</td></tr>
<tr><td colspan="6">1340 cm$^{-1}$, medium</td><td>Ph-N stretching for quaternary ammonium</td></tr>
<tr><td colspan="6">1420 cm$^{-1}$, weak to very weak</td><td>$CH_2$ scissoring</td></tr>
<tr><td colspan="6">1460 cm$^{-1}$,  medium</td><td>semicircle stretching mixed with a C-H bending</td></tr>
<tr><td colspan="6">1600 cm$^{-1}$, medium</td><td>Ring stretching for dimethyl aniline</td></tr>
<tr><td colspan="5">2925/2970 cm$^{-1}$, strong</td><td>--</td><td>$CH_2$ asymmetric stretching</td></tr>
<tr><td colspan="2">3070 cm$^{-1}$, weak</td><td colspan="4">--</td><td>Vinyl $CH_2$ stretching</td></tr>
</table>

TABLE 1 Position, assignment and intensity of PEG, SPS and BPPEI bands

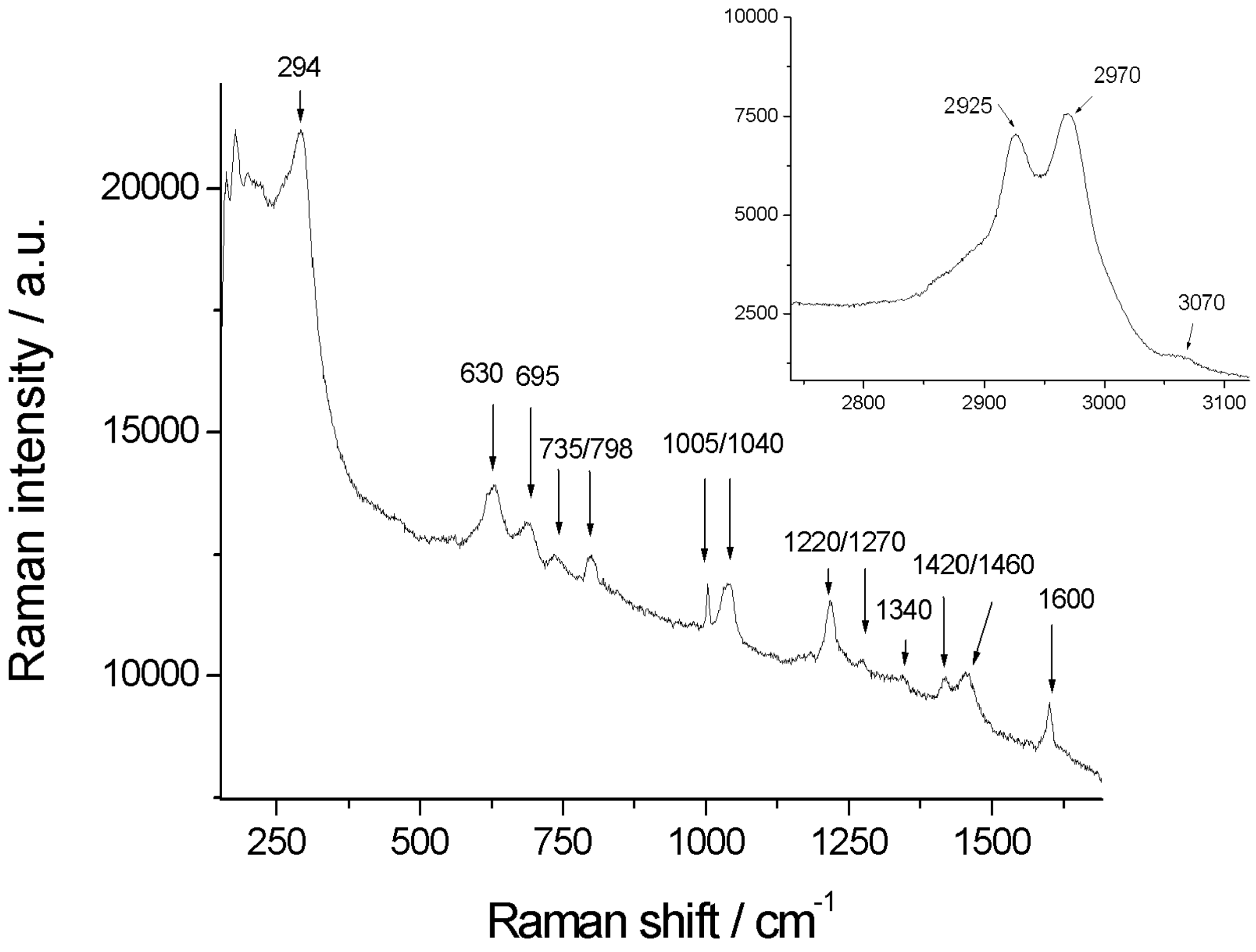

Figure 1 - In situ SERS spectra recorded at OC from a bath, containing 50 ppm $Cl^-$, 300 ppm PEG, 6 ppm SPS, and 1 mg $L^{-1}$ BPPEI.

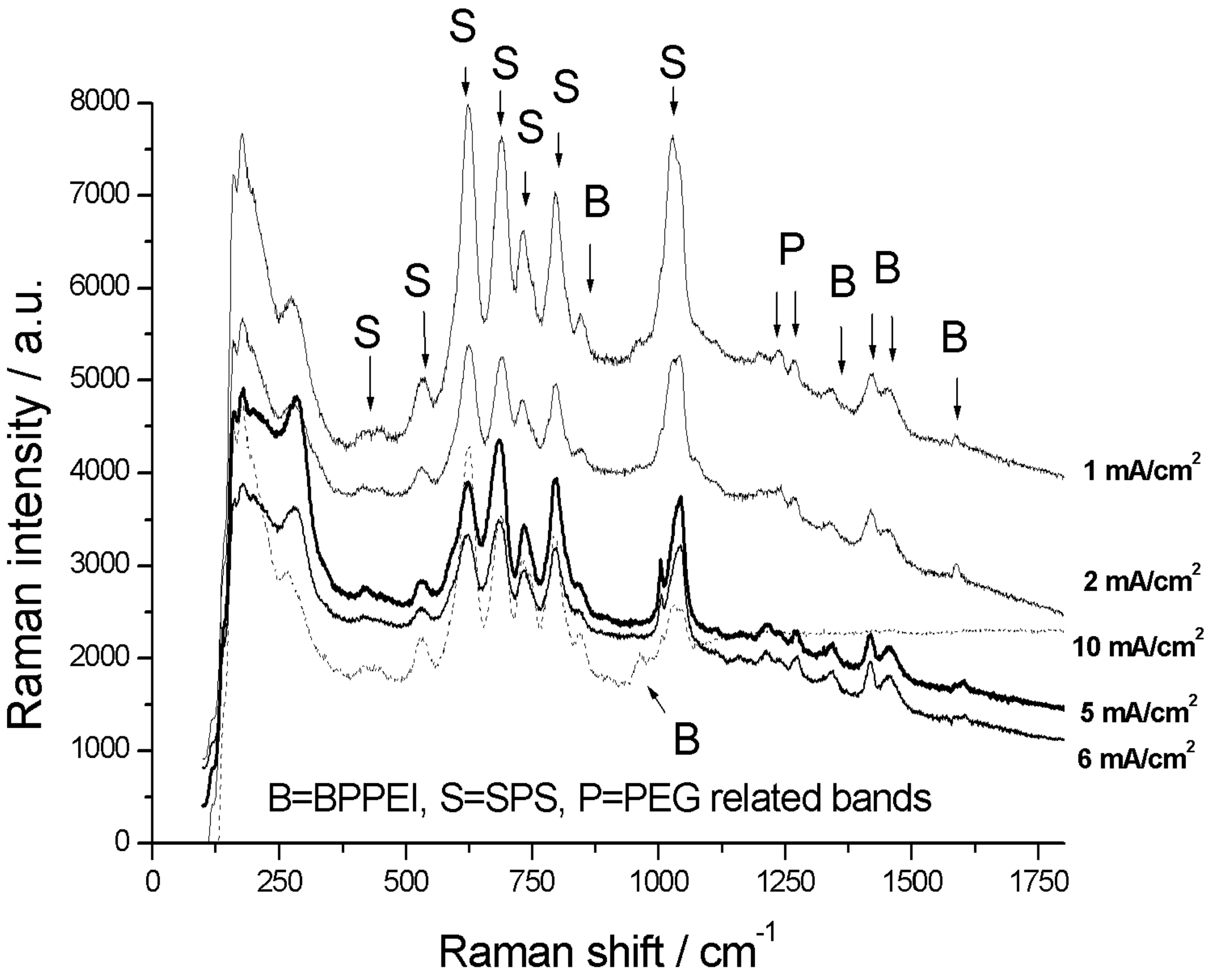

Figure 2 - In situ SERS spectra recorded from a bath, containing 50 ppm $Cl^-$, 300 ppm PEG, 6 ppm SPS, and 1 mg $L^{-1}$ BPPEI at the indicated current densities.

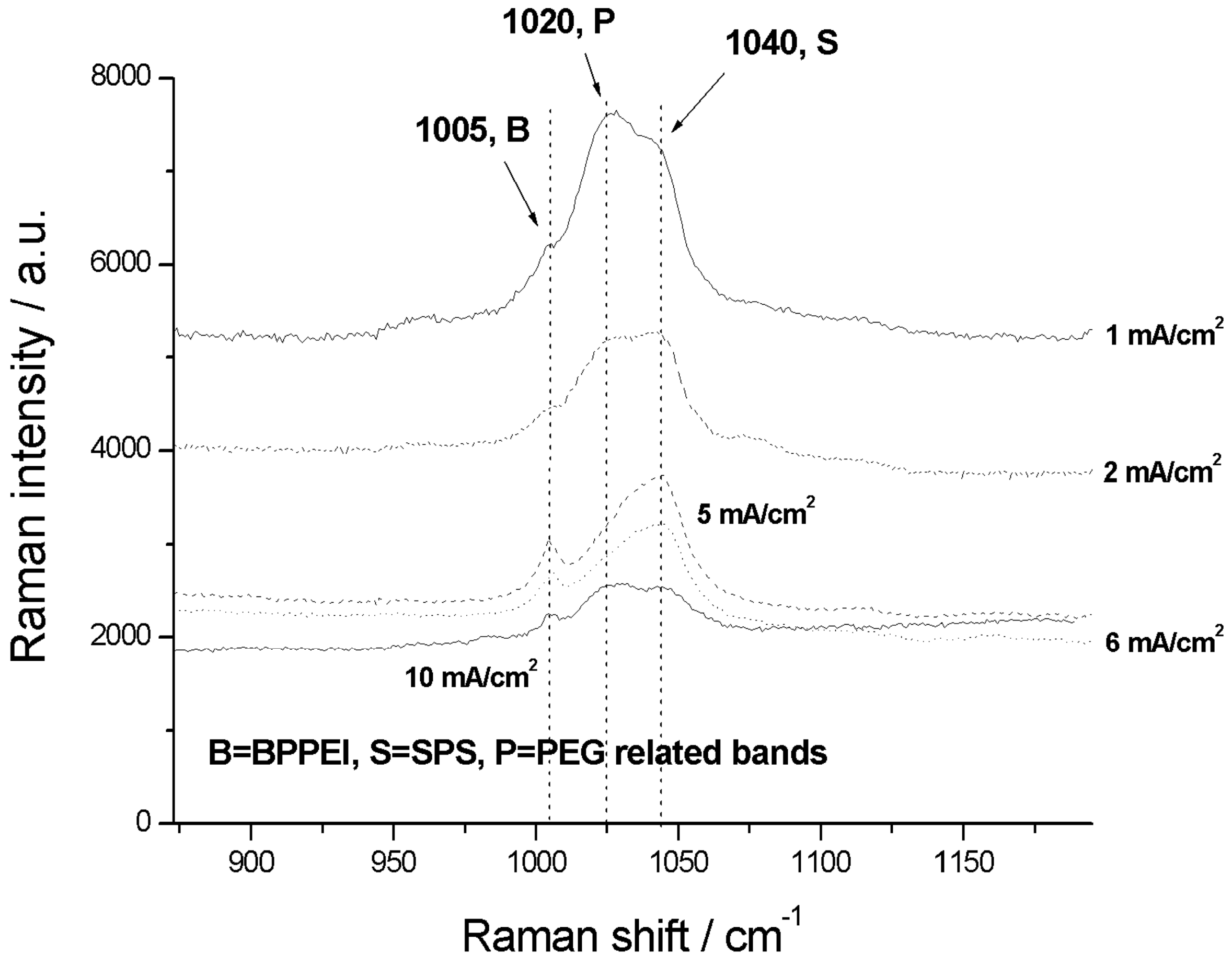

Figure 3 - In situ SERS spectra (range 900÷1250 $cm^{-1}$) recorded from a bath containing 50 ppm $Cl^{-}$, 300 ppm PEG, 6 ppm SPS, and 1 mg $L^{-1}$ BPPEI at the indicated current densities.

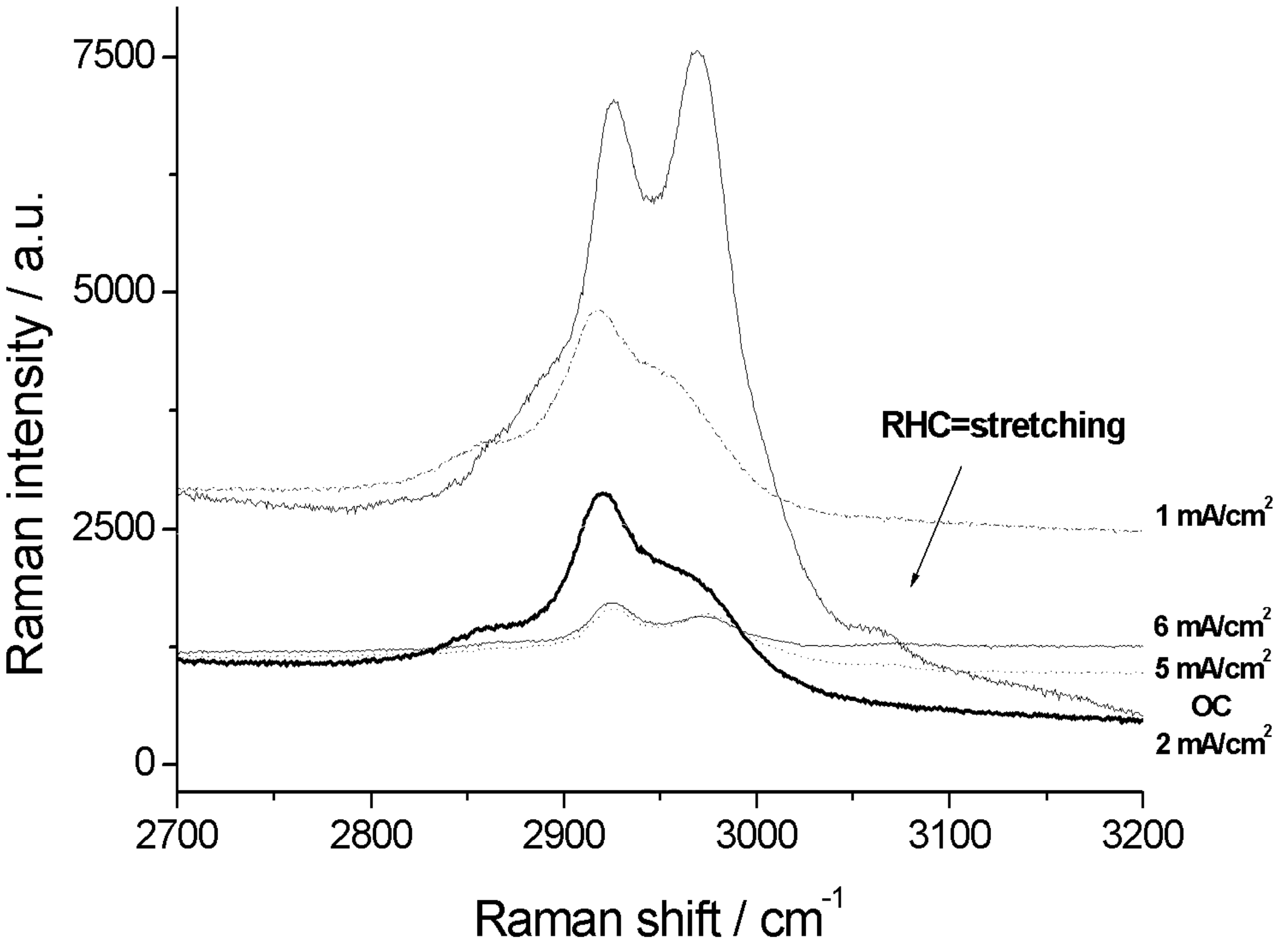

Figure 4 - In situ SERS spectra (range 2700÷3200 $cm^{-1}$) recorded from a bath containing 50 ppm $Cl^{-}$, 300 ppm PEG, 6 ppm SPS, and 1 mg $L^{-1}$ BPPEI at the indicated current densities.

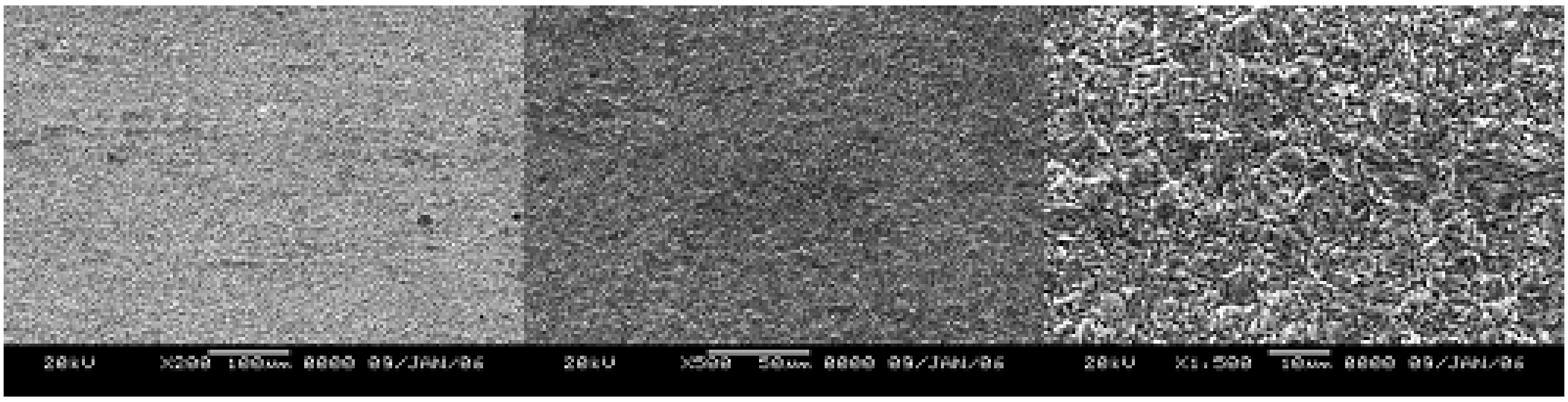

Figure 5 - SEM micrographs of Cu deposited from a solution containing $CuSO_4 \cdot 5H_2O$ 20mM, $H_2SO_4$ 0.5M, NaCl 50ppm, PEG MW 1500, 300 ppm, SPS 6 ppm and 3-Diethylamino-7-(4-dimethylaminophenylazo)-5-phenylphenazinium chloride (Janus Green B, JGB) 1 mg $L^{-1}$, obtained at 2 mA $cm^{-2}$, magnification 200x, 500×, and 1500x.

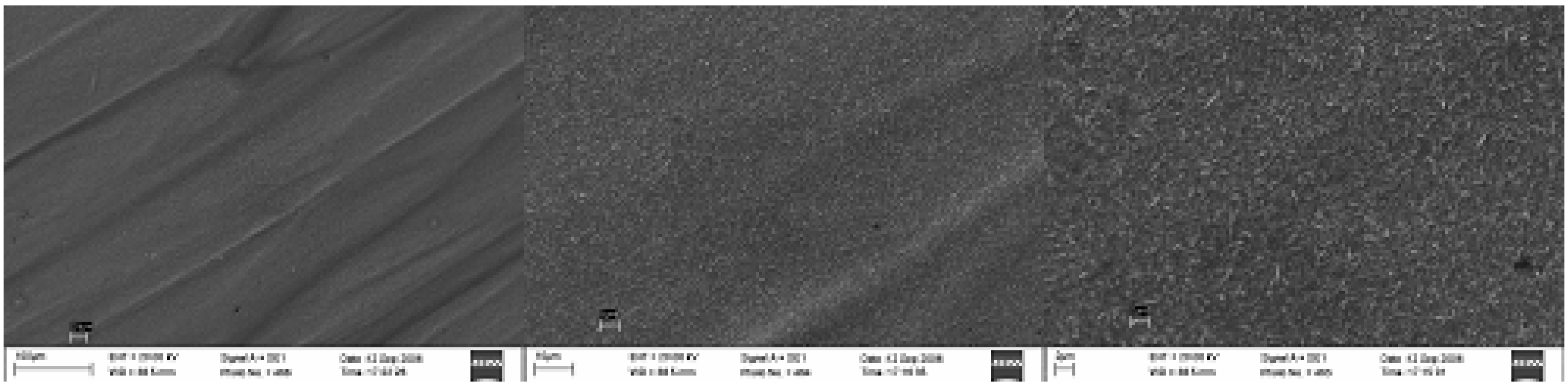

Figure 6 - SEM micrographs of Cu deposited from a solution containing $CuSO_4 \cdot 5H_2O$ 20mM, $H_2SO_4$ 0.5M, NaCl 50ppm, PEG MW 1500, 300 ppm, SPS 6 ppm and BPPEI 10 mg $L^{-1}$, obtained at 2 mA $cm^{-2}$, magnification: 500×, 1500x, and 6000x.